# Phosphorus Spin Coherence Times in Silicon at Very Low Temperatures

L.K. Alexander, N. Suwuntanasarn, W.D. Hutchison

*ARC Centre for Quantum Computer Technology,
School of Physical, Environmental and Mathematical Sciences,
The University of New South Wales, ADFA, Canberra, ACT 2600, Australia.*

*Phosphorus donor spin coherence in isotopically pure 28 silicon is measured at very low temperatures using pulsed electron spin resonance. The isolated spin $T_2$ varies unexpectedly with phosphorus concentration.*

Keywords: magnetic resonance, donor spins in silicon, quantum computing

## I. INTRODUCTION

Phosphorus donor atoms ($^{31}$P) in silicon (Si:P) are expected to have very long (both nuclear and electron) spin relaxation times at low temperature. This combined with the obvious compatibility of silicon with existing device fabrication technology, makes this system of interest as a potential basis for quantum computing (QC) [1]. In Si:P, the dephasing of the donor electron spin represents the decoherence time of the device (single qubit decoherence). Pulsed electron spin resonance (ESR) offers a convenient and most effective way to study this dephasing. The original pulsed ESR studies of Si:P were conducted decades ago [2][3]. However, since the renewed interest in Si:P for QC applications, further work has been done. In particular, a projected isolated spin decoherence time ($T_2$) of 60 ms in an epilayer of isotopically enriched $^{28}$Si at 7 K was reported more recently [4]. There is considerable potential for this number to be improved (i.e. lengthened). In particular improvements in the $^{28}$Si purity are important since secondary hyperfine interactions between the donor spin and nuclear spin of I = ½ $^{29}$Si nuclei are a significant source of line broadening and spin decoherence. Also it might be expected that reduced phosphorus concentrations are desirable in the quest to establish $T_2$ in the single spin limit when using an ensemble measurement. There is an intrinsic decoherence caused by ensemble rotation of the refocusing pulse in an electron spin echo sequence and although this effect, also known as instantaneous diffusion (ID), may in principle be removed though projection to zero second pulse turn angle, there is a limit to the practical deconvolution of components with vastly different time scales. Finally lowering the temperature also has potential. The spin lattice relaxation rate ($T_1$), which provides the upper limit on $T_2$, get extremely long in Si:P at low temperature. Not only does the spin lattice relaxation ($T_1$), which provides the upper limit on $T_2$, get extremely long in Si:P at low temperature, but recently it has been suggested that decoherence based on pairwise interactions such as dipolar interactions can be suppressed at very low temperatures [5].

Here we report on some Si:P decoherence time measurements using two different $^{28}$Si enriched samples, an epilayer with doping of 1 x 10$^{16}$ P cm$^{-3}$ and a bulk wafer with doping of 5 x 10$^{15}$ P cm$^{-3}$ from which we estimate isolated spin $T_2$'s for phosphorous donor electron spins.

## II. EXPERIMENTAL DETAILS

The $^{28}$Si:P epilayer samples were from ISONICS and are 10 μm thick on a Si[100] highly resistive p-type, 100 μm thick substrate. The phosphorus doping concentration in the epilayer is approximately 1.0 x 10$^{16}$ P cm$^{-3}$ and the $^{29}$Si fraction is below 0.1%. The other sample for which results are discussed below is a piece of bulk, phosphorus doped, isotopically enriched $^{28}$Si ($^{28}$Si:P) with a concentration ~5 x 10$^{15}$ P cm$^{-3}$ and purity 99.92 %. In both the samples the doping concentrations were determined by Hall bar measurements.

Electron spin echo (ESE) pulse sequences were performed with the typical (π/2-τ-π-τ-echo) set at (16ns-τ-32ns-τ-echo). The resonant frequency used was approximately 9.4 GHz, and we tune the system and magnetic field to resonate the higher field (lower g factor) resonance satellite of the Si:P hyperfine split doublet since this line is clear of any extraneous surface charge trap resonance lines. With our set-up, ESE can be carried out at temperatures down to 60 mK. However, to sensibly follow the trends in the coherence time and allow the use of light to reset the spins between pulse sequences, ESE measurements were carried out at 4.2 K and 0.9 K. Our preliminary results at 0.2 K is reported in ref [6].

As mentioned above, the spin lattice relaxation time ($T_1$) for Si:P becomes very long at low temperatures. For P concentrations below ~10$^{16}$ P cm$^{-3}$ the rate varies with a T$^7$ power down to 2 K and continues at T$^1$ below 2 K, with $T_1$ reaching >1000 s at 1.2 K [7]. This represents a major obstacle for signal averaging in ESE experiments since a delay of ~5 times $T_1$ should be applied between each pulse sequence. However reports (eg [7]) have shown that $T_1$ could be shortened dramatically by the application of light with photon energy above the bandgap (> ~1.0 μm). In this work we used 1 s wide bursts (20 mW of 532 nm green laser light directed down a plastic light guide), followed by a 60 s wait time, between each pulse sequence. This was based on a comprehensive study of the effect of light performed in our earlier work[8,9]. Such a laser light sequence generated a significant shortening of the relaxation time at temperatures down to 0.9 K, but no effect on the resulting shape of the echo decay curves as compared to waiting for much longer times between pulse trains.

The Si:P echo decay results were fitted using the following expression:

$$V(\tau) = V_0 \exp[-(2\tau/T_M) - (2\tau/T_{SD})^n] \quad (1)$$

where $T_M$ is the ensemble exponential decay time constant, which incorporates several terms detailed below, $T_{SD}$ is the spectral diffusion (SD) decay time associated mainly with the interaction with $^{29}Si$ nuclei and $n$ is an exponential stretching factor varied between 2 and 3 for different SD regimes by different authors[2,10]. The intrinsic phase memory time, $T_2$, also defined as decoherence time of an isolated electron spin free from the effect of ID, is derived from $T_M$ and $T_{ID}$ via the relationship $1/T_M = 1/T_2 + 1/T_{ID}$. Here $T_{ID}$ may be estimated from $1/T_{ID} = C\pi\mu g^2 \mu_B^2 \sin^2(\beta/2)/(9\sqrt{3}\hbar)$, where $C$ is the concentration of the excited electron spins (for the concentration P in the sample, [P] = 2C), $\mu$ is the permeability of crystalline silicon, $g$ is the g-factor of the donor electron, $\mu_B$ is the Bohr magneton and $\beta$ is the turn angle of the refocusing pulse.

In the case of the isotopically pure samples used here, we find that the $T_{SD}$ term can be largely ignored based on the fact that the echo decays can be fitted well to obtain estimates of $T_M$ with a single exponential function. Therefore the Eqn (1) modifies as:

$$V(\tau) = V_0 \exp[-(2\tau/T_M)] \quad (2)$$

To estimate $T_2$, we use the approach used by [4] in Si:P, and originally developed by [11] where it was recognised that since $T_{ID}$ is proportional $\sin^2(\beta/2)$, it is better to carry out a series of experiments with different values of β and then project to β = 0 to find the value of $T_2$ rather than relying on a multi-parameter fit of a single data set.

III. RESULTS AND DISCUSSION

A. *1 x 10$^{16}$ P cm$^{-3}$ Epilayer sample*

For the epilayer sample, ESE measurements were carried out at temperatures of 7.0 K, 4.2 K and 1.0 K. It is to be noted that light resetting of the spins was only required for 4.2 K and 1.0 K. From Figure 1, it can be observed that the echo decay and thereby decoherence time increases as the temperature is decreased regardless of the type of fitting method applied. All the data shown in Figure 1 was collected with the second pulse width corresponding to π or 180 degrees. The echo decay trend for $^{28}Si:P$ is almost single exponential, indicating little influence from a SD term. To test this premise these results were fitted both with and with out a SD term. In the case with an SD term, n is fixed at 3 following [4].

The $T_2$ values found by fitting Equation (1) with $T_{SD}$ were 6.3 ± 0.5 ms, 7.0 ± 0.5 ms and 8.2 ± 0.7 ms respectively at 7.0 K, 4.2 K and 1.0 K. The values for $T_2$ neglecting the $T_{SD}$ factor, fitting equation 2 were similar to the values estimated from fitting using equation 1. The values were 6.3 ± 0.5 ms, 6.8 ± 0.5 ms and 9.3 ± 0.7 ms respectively at 7.0 K, 4.2 K and 1.0 K. In fact we are inclined to adopt these latter values since method in equation 2 does not require the somewhat subjective assignment of $T_{SD}$ and n.

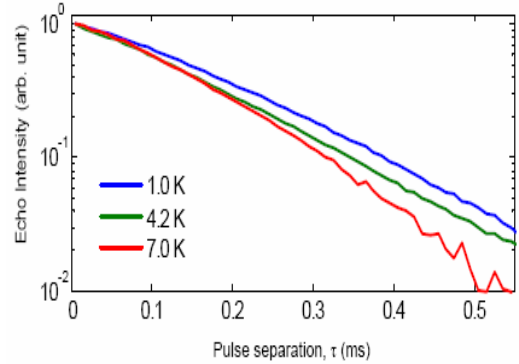

Figure 1. The echo decay of $^{28}Si:P$ epilayers with P concentration of 1x10$^{16}$ P/cm$^3$ at different temperatures.

Our $T_M$ value at 7.0 K (0.31 ms) is in good agreement with that of the sample of comparable concentration and at the same temperature in the study of [4]. Also it can be noted that in the range 7.0 K to 1.0 K, the decoherence time $T_2$ increases with the reduction in temperature. This result is in contrast to the report [4] on their experiments at 7-20 K. In their case, compared to 8.1 K, there was only marginal improvement in $T_2$ values when the temperature was brought down to 6.9 K. Note that the $T_1$ relaxation processes dominant in the temperature range of our investigations are Raman (4.2 K) and direct phonon (1.0 K) as compared to Orbach relaxation in the higher temperature regime. $T_1$ is of course providing an upper bound for $T_2$.

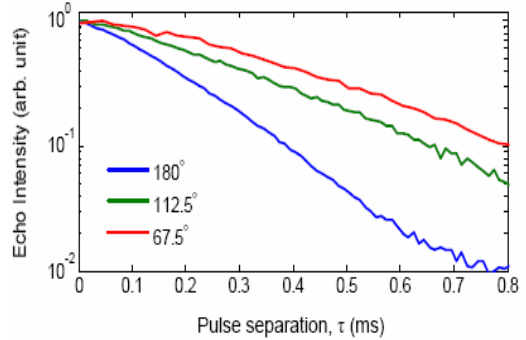

Figure 2. The echo decay of $^{28}Si:P$ epilayers with P concentration of 1x10$^{16}$ P/cm$^3$ at different second pulse turn angles, measured at T = 1.0 K.

In Figure 2 the variation in the echo decay rate at 1.0 K for different second pulse turning angles is demonstrated. As noted earlier the expected variation of $T_M$ with second pulse turn angle β can be used to provide a more satisfactory way to estimate $T_2$. This method is applied to 1.0 K ESE data for the epilayer sample in Figure 3. The projection to zero second pulse turn angle resulting in a value for $T_2$ of 10 ± 4 ms in agreement with the earlier estimate for this sample at 1 K.

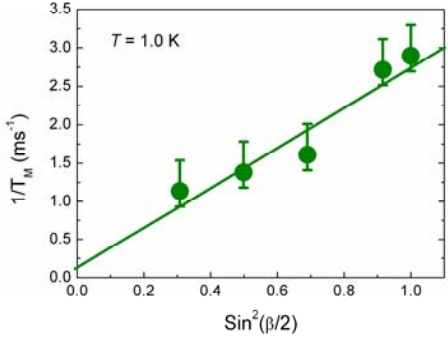

Figure 3. The extrapolation of Si:P $1\times10^{16}$ P/cm$^3$ echo decay for different second pulse turn angles at temperature, $T = 1.0$ K.

## B.  $5\times10^{15}$ P cm$^{-3}$ bulk $^{28}$Si:P sample

Echo decay data collected for the bulk $^{28}$Si:P sample at temperatures of 4.2 K and 0.9 K with various second pulse turn angles are shown in Figures 4 and 5 respectively. The resulting ensemble $T_M$ values are also listed on the figures. There is a trend apparent that the $T_M$ values for 0.9 K are longer than at 4.2 K for the same value of β. This data is then used to generate the plots of $1/T_M$ v $\sin^2(\beta/2)$ at 4.2 K and 0.9 K of Figure 6. In this figure, some of our earlier data [6] is also added. Comparison of the two data sets at 4.2 K shows the repeatability with in the estimated errors. As before, linear fits extending to β= 0 are used to estimate the single spin limit decoherence time, $T_2$. We obtain $T_2 = 260$ (50) ms at 4.2 K and 330 (100) ms at 0.9 K for the bulk sample.

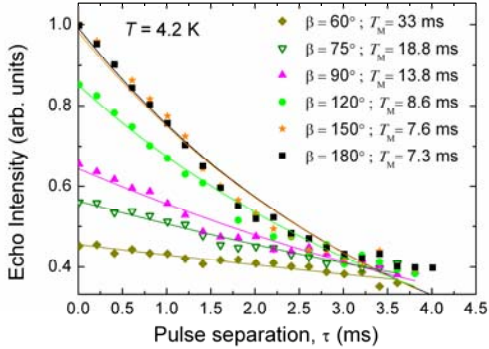

Figure 4. The echo decay of $^{28}$Si:P bulk samples with P concentration $5\times10^{15}$ P cm$^{-3}$ at different refocusing pulse turn angles (β) at $T = 4.2$ K.

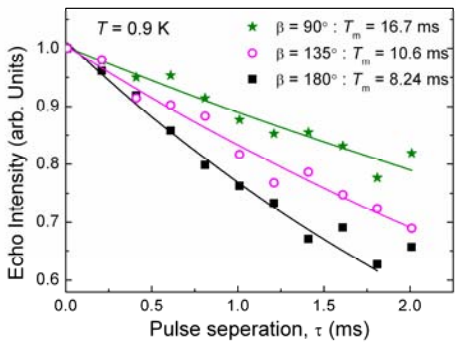

Figure 5. The echo decay of $^{28}$Si:P bulk samples with P concentration $5\times10^{15}$ P cm$^{-3}$ at different refocusing pulse turn angles (β) at $T = 0.9$ K

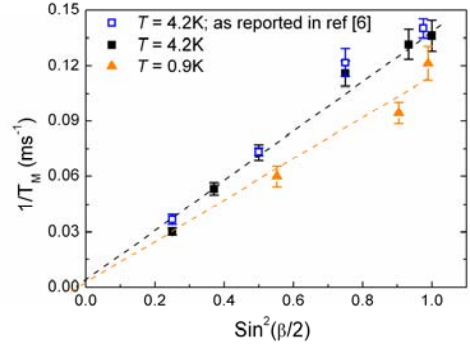

Figure 6: A plot of the relaxation rate ($1/T_M$) as a function of the refocusing pulse turn angles. The line is extrapolated to small β to obtain the single spin relaxation rate, $1/T_2$.

## IV.  CONCLUSION

In this paper we have presented ESE data collected from two isotopically ennriched and phosphorus doped $^{28}$Si:P samples: an epilayer with doping of $1 \times 10^{16}$ P cm$^{-3}$ and a bulk wafer with doping of $5 \times 10^{15}$ P cm$^{-3}$. From this ESE data we have estimated isolated spin $T_2$'s for phosphorous donor electron spins using a projection to zero second pulse tip angle to remove the effect of instantaneous diffusion (ID). We find that for both samples $T_2$ increases as the temperature is lowered towards 1 K.

The adopted values for $T_2$ at 4.2 K were ~7 ms and ~260 ms for the epilayer sample and bulk sample, respectively. While at ~1 K values of 10 ms and 330 ms respectively were also estimated. There is therefore a 30+ fold increase in the value of low temperature decoherence time between the two samples. This is somewhat unexpected since at low temperature $T_2$ is thought to be mainly dictated by the dipole – dipole interactions between the donor electron spins and/or the residual $^{29}$Si nuclei. Here the $^{29}$Si concentration is nominally less than 0.1% for both samples, i.e. not very different, and also there is only a nominal two fold reduction in phosphorus concentration for the bulk wafer. Of course we should not discount the possibility that some of the difference is due to the nature of the epilayer sample. Anecdotally such epilayers can contain additional strain and crystalline imperfection, and might also have significant non-zero spin impurities (other than P). Alternatively it has been suggested by several authors that what is important for a long coherence time in the presence of dipolar coupling is a low total effective dipolar field, summed over all sites, at the target donor site (e.g. see [12]) and it is conceivable that this could occur at a particular concentration that is not simply the lowest possible.

What ever the case, we have observed decoherence times of several hundred milliseconds in $^{28}$Si:P and it seems quite likely that this value for genuinely isolated nuclei could be in excess of one second. Definitely the system is attractive for quantum computing applications.

ACKNOWLEDGMENT

The authors wish to thank Prof. M.S. Brandt for the loan of the bulk sample. This work is supported by the Australian Research Council and by The University of New South Wales.